\documentclass[traditabstract,letter]{aa}
\usepackage{graphicx}
\usepackage{natbib}
\usepackage{epstopdf}
\bibpunct{(}{)}{;}{a}{}{,}
\usepackage{txfonts}

\begin{document}
\title{
Constraints on mass loss and self-enrichment scenarios for the globular clusters of the Fornax dSph 
\thanks{
Based on observations made with ESO Telescopes at the La Silla Paranal Observatory under programme ID 078.B-0631(A)
}
}

\author{S. S. Larsen \inst{1} \and
  J. Strader \inst{2,3} \and
  J. P. Brodie \inst{4}
}
\institute{Department of Astrophysics / IMAPP, Radboud University Nijmegen, P.O. Box 9010, 6500 GL Nijmegen, The Netherlands
\and
  Harvard-Smithsonian Center for Astrophysics, 60 Garden Street, Cambridge, MA 02138, USA
\and
  Department of Physics and Astronomy, Michigan State University, East Lansing, Michigan 48824, USA
\and
  UCO/Lick Observatory, University of California, Santa Cruz, CA 95064, USA
}

\offprints{S.\ S.\ Larsen, \email{s.larsen@astro.ru.nl}}

\date{Received 26 June 2012 / Accepted 19 July 2012}

\abstract
{
 Recently, high-dispersion spectroscopy has demonstrated conclusively that four of the five globular clusters (GCs) in the Fornax dwarf spheroidal galaxy are very metal-poor with ${\rm [Fe/H]}<-2$. The remaining cluster, Fornax 4, has ${\rm [Fe/H]}=-1.4$. This is in stark contrast to the field star metallicity distribution which shows a broad peak around ${\rm [Fe/H]}\approx-1$ with only a few percent of the stars having ${\rm [Fe/H]}<-2$. 
If we only consider stars and clusters with ${\rm [Fe/H]}<-2$ we thus find an extremely high GC specific frequency, $S_{\! N}\approx400$, implying by far the highest ratio of GCs to field stars known anywhere. We estimate that about 1/5--1/4 of all stars in the Fornax dSph with  ${\rm [Fe/H]}<-2$ belong to the four most metal-poor GCs. These GCs could, therefore, at most have been a factor of 4--5 more massive initially.
Yet, the Fornax GCs appear to share the same anomalous chemical abundance patterns known from Milky Way GCs, commonly attributed to the presence of multiple stellar generations within the clusters. The extreme ratio of metal-poor GC- versus field stars in the Fornax dSph is difficult to reconcile with scenarios for self-enrichment and early evolution of GCs in which a large fraction (90\%--95\%) of the first-generation stars have been lost. 
It also suggests that the GCs may not have formed as part of a larger population of now disrupted clusters with an initial power-law mass distribution. The Fornax dSph may be
a rosetta stone for constraining theories of the formation, self-enrichment and early dynamical evolution of star clusters.
}


\titlerunning{Globular clusters in the Fornax dSph}
\maketitle

\section{Introduction}

The Fornax dwarf spheroidal galaxy (dSph) is well-known for its high globular cluster (GC) specific frequency, i.e., the number of GCs normalized to a host galaxy $M_V=-15$  \citep{Harris1981}. Assuming $M_V = -13.2$ \citep{Mateo1998}, the five GCs \citep{Hodge1961} correspond to a specific frequency of $S_{\! N}=26$.
Specific frequencies rivalling that of Fornax have only been found in other similarly faint dwarf galaxies \citep{Peng2008,Georgiev2010}.
Spiral galaxies typically have $S_{\! N}\approx 1$, while normal elliptical galaxies  have $S_{\! N}\approx3-5$, slightly higher in clusters than in the field \citep{Harris1991}. Even the most cluster-rich cD galaxies generally have $S_{\! N}\la15$ \citep{Harris1991,Brodie2006}. 
These differences constitute the classical ``specific frequency problem''.

A second specific frequency problem becomes apparent when considering the numbers of GCs associated with different stellar populations \emph{within} galaxies. Early-type galaxies are generally redder (more metal-rich) than the average  of their GC systems \citep{Forte1981,Larsen2001,Forbes2001}, and direct comparisons of metallicity distributions for stars and globular clusters have confirmed that the specific frequency tends to increase with decreasing metallicity within galaxies \citep{Harris2002,Harris2007}. This is also evident in our own Galaxy: of the roughly 150 GCs known in the Milky Way, about 2/3 are associated with the halo \citep{Zinn1985}, even though the halo only contains 1\%--2\% of the stellar mass \citep{Suntzeff1991,Dehnen1998}.

It has, in fact, been suggested that a significant fraction of the Milky Way halo stars originate from disrupted GCs. If the Galactic GC population  formed with a power-law mass function similar to that observed in young cluster systems at the present epoch \citep{Larsen2009}, then the stars lost from dissolving  clusters over a Hubble time might be sufficient to account for essentially the entire stellar halo \citep[e.g.][]{Kruijssen2009}.

The above estimate only involves standard dynamical evolution of the clusters. However, a similar conclusion can be reached via a different line of reasoning, starting from the increasing body of evidence that GCs host dual or multiple stellar generations \citep{Gratton2012}. In the following we will refer to the ``first'' and ``second'' generation of stars, keeping in mind that the actual star formation histories may be more complex.
The second generation is characterized, among other things, by anomalous abundances of several light elements that imply $p$-capture nucleosynthesis at high temperatures. 
The currently favoured sites are massive AGB stars \citep{DErcole2008}, fast-rotating massive main sequence stars \citep{Decressin2007}, or massive interacting binaries  \citep{deMink2009}.
A fundamental challenge faced by most scenarios for the origin of multiple stellar generations in GCs is the ``mass budget'' problem: at the present epoch, the second generation of stars has a mass similar to, or even greater than the first generation. However, for a standard IMF the ejecta produced by the first-generation stars fall short by large factors  compared to the mass required to form the observed second generation, even if a 100\% star formation efficiency is assumed. This has led to the suggestion that a large fraction of the first generation, up to 90--95\%, was lost soon after the formation of the second generation
\citep{DErcole2008,Schaerer2011,Bekki2011}. Accounting for subsequent dynamical evolution, one again finds that a significant fraction of the Milky Way stellar halo might come from GCs, even without considering that a significantly larger population of lower-mass clusters may have been present initially \citep{Gratton2012}.

In this letter we discuss the globular cluster system of the Fornax dSph in the context of the second specific frequency problem and self-enrichment scenarios. While these two issues may seem unrelated, the Fornax dSph turns out to be such an extreme case of the second $S_{\! N}$ problem that it puts tight constraints on the amount of mass that could have been lost from its GCs. 

\section{Chemical composition of the Fornax GCs}

\begin{table}
\begin{minipage}[t]{\columnwidth}
\caption{Metallicities and absolute magnitudes for the Fornax GCs.
}
\label{tab:fornaxgc}
\renewcommand{\footnoterule}{}
\begin{tabular}{lcc} \hline\hline
                 & [Fe/H]           &   $M_V$ \\ \hline
Fornax 1 & $-2.5\pm0.1^1$    & $-5.2^3$   \\
Fornax 2 & $-2.1\pm0.1^1$    & $-7.3^3$   \\
Fornax 3 & $-2.3\pm0.1^2$    & $-8.2^3$ \\
Fornax 4 & $-1.4\pm0.1^2$    & $-7.2^3$ \\
Fornax 5 & $-2.1\pm0.1^2$    & $-7.4^3$  \\
\hline
\end{tabular}
\tablebib{
(1)~\citet{Letarte2006}; (2)~\citet{Larsen2012}; (3)~\citet{Webbink1985}
}
\end{minipage}
\end{table}

We have recently carried out a detailed analysis of the chemical composition of the three clusters Fornax~3, 4 and 5 from integrated-light high-dispersion spectroscopy with UVES on the ESO Very Large Telescope. Our analysis techniques are described in a companion paper \citep[][submitted]{Larsen2012},
along with a more detailed discussion of the abundance measurements and references to previous determinations of the metallicities.
We measured iron abundances of ${\rm [Fe/H]} = -2.3\pm0.1$ (Fornax 3), $-1.4\pm0.1$ (Fornax~4) and $-2.1\pm0.1$ (Fornax~5). The clusters Fornax~1, 2 and 3 were observed by \citet{Letarte2006}, who measured abundances for three individual RGB stars in each cluster; they obtained ${\rm [Fe/H]} = -2.5\pm0.1$ (Fornax 1), $-2.1\pm0.1$ (Fornax~2) and $-2.4\pm0.1$ (Fornax~3). We note that there is excellent agreement between the two independent measurements for Fornax~3. 
The high-dispersion measurements of the Fe abundances are summarized in Table~\ref{tab:fornaxgc}. It is clear that Fornax~4 is significantly more metal-rich than the other clusters, all of which have ${\rm [Fe/H]}<-2$. 
We also list the GC $M_V$ magnitudes, based on the magnitudes in \citet{Webbink1985} and using the same apparent distance modulus of $m-M=20.80$ as for the Fornax dSph itself for consistency  \citep{Mateo1998,Pietrzynski2009}.

Data on abundance variations within the clusters are still relatively limited.  \citet{Letarte2006} noted that two of their nine stars show enhanced Na and depleted O abundances, as well as low Mg abundances. These patterns are similar to those observed in putative second-generation stars in Milky Way globular clusters. In our integrated-light measurements we  observed Mg to be depleted relative to other $\alpha$-elements (Ca, Ti) in Fornax~3, 4 and 5, again suggesting that some stars in these clusters display anomalous abundances. It thus appears that the Fornax GCs are similar to the Milky Way GCs in this respect.

\section{The globular cluster specific frequency revisited}
\label{sec:sn}

\begin{figure}
\centering
\includegraphics[width=\columnwidth]{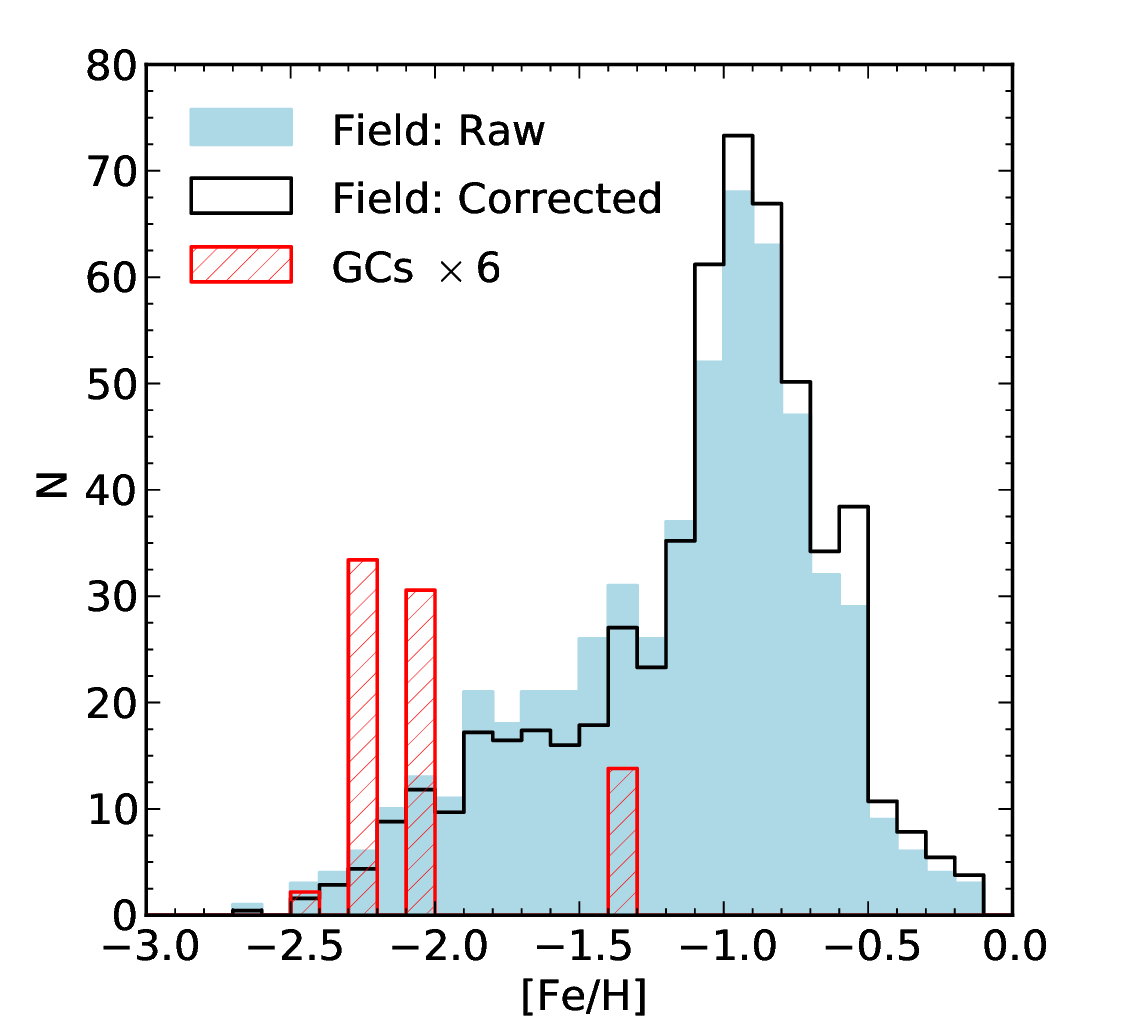}
\caption{\label{fig:zglob}Metallicity distributions for field stars and globular clusters in the Fornax dSph. The filled histogram shows the raw distribution of field star metallicities from \citet{Battaglia2006}, while the outlined histogram is corrected for radial variations in the coverage (see main text). The hashed histogram represents the GCs with weights scaled by their luminosities and exaggerated by a factor of 6 (note that Fornax~2 and Fornax~5 fall in the same bin).}
\end{figure}

\citet{Battaglia2006} measured metallicities for 562 field red giants in the Fornax dSph covering the full radial range out to  $\sim1\fdg2$, using spectroscopy of the \ion{Ca}{ii} near-infrared triplet obtained with the FLAMES spectrograph on the ESO VLT. 
About 7\% of the stars in their sample have metallicities ${\rm [Fe/H]}<-2$. However, the raw sample may not accurately represent the true metallicity distribution of the Fornax stars as the metallicity distribution does show a radial variation (with more metal-rich stars in the centre), and no correction has been made for the varying degree of completeness as a function of galactocentric distance. We have attempted to make a rough correction for this effect, proceeding as follows:

We divided the  profile of the Fornax dSph into 10 radial bins. For each bin, we computed the total number of stars $N_T(R)$ from the Sersic model fit of \citet{Battaglia2006} and counted the number of stars with FLAMES metallicity measurements, $N_F(R)$. We then assigned a weight, $W(R) = N_T(R) / N_F(R)$, to each star in the corresponding bin. Finally, a corrected metallicity distribution was derived by counting each star according to its weight. The raw and corrected metallicity distributions are shown in Fig.~\ref{fig:zglob}.
Perhaps somewhat surprisingly, this procedure leads to a somewhat \emph{less} pronounced low-metallicity tail, with about 5\% of the stars having ${\rm [Fe/H]}<-2$. 
This is because a much smaller fraction of the stars within a given FLAMES field are actually sampled in the central regions of the galaxy, so that the weights $W(R)$ increase towards the centre where there are fewer very metal-poor stars.
For comparison we also include the GCs in Fig.~\ref{fig:zglob}. Each cluster was ``counted'' as $6 \times N_\mathrm{stars} \times 10^{-0.4 (M_{V,\rm GC} + 13.2)}$ stars, where $N_\mathrm{stars} = 562$ is the total number of stars with FLAMES data, $M_{V,\rm GC}$ are the $M_V$ magnitudes of the individual clusters and $M_V=-13.2$ is the magnitude of the Fornax dSph. The difference between the metallicity distributions of the field stars and the GCs is rather striking.

If we simply scale the luminosity of the Fornax dSph by 5\%, we obtain a total absolute magnitude of $M_V=-10.0$ for the low-metallicity tail. Relative to this tail, the four metal-poor GCs then correspond to a specific frequency of $S_{\! N} = 400$. With an absolute magnitude of  $M_V=-8.9$ these GCs account for over 1/4 of the integrated luminosity of all the stars with ${\rm [Fe/H]} <-2$ in Fornax!  
Note that this calculation assumes that numbers of RGB stars and integrated luminosities are equivalent, which is only approximately true as the number of RGB stars per unit luminosity depends somewhat on age and metallicity.

\begin{figure}
\centering
\includegraphics[width=85mm]{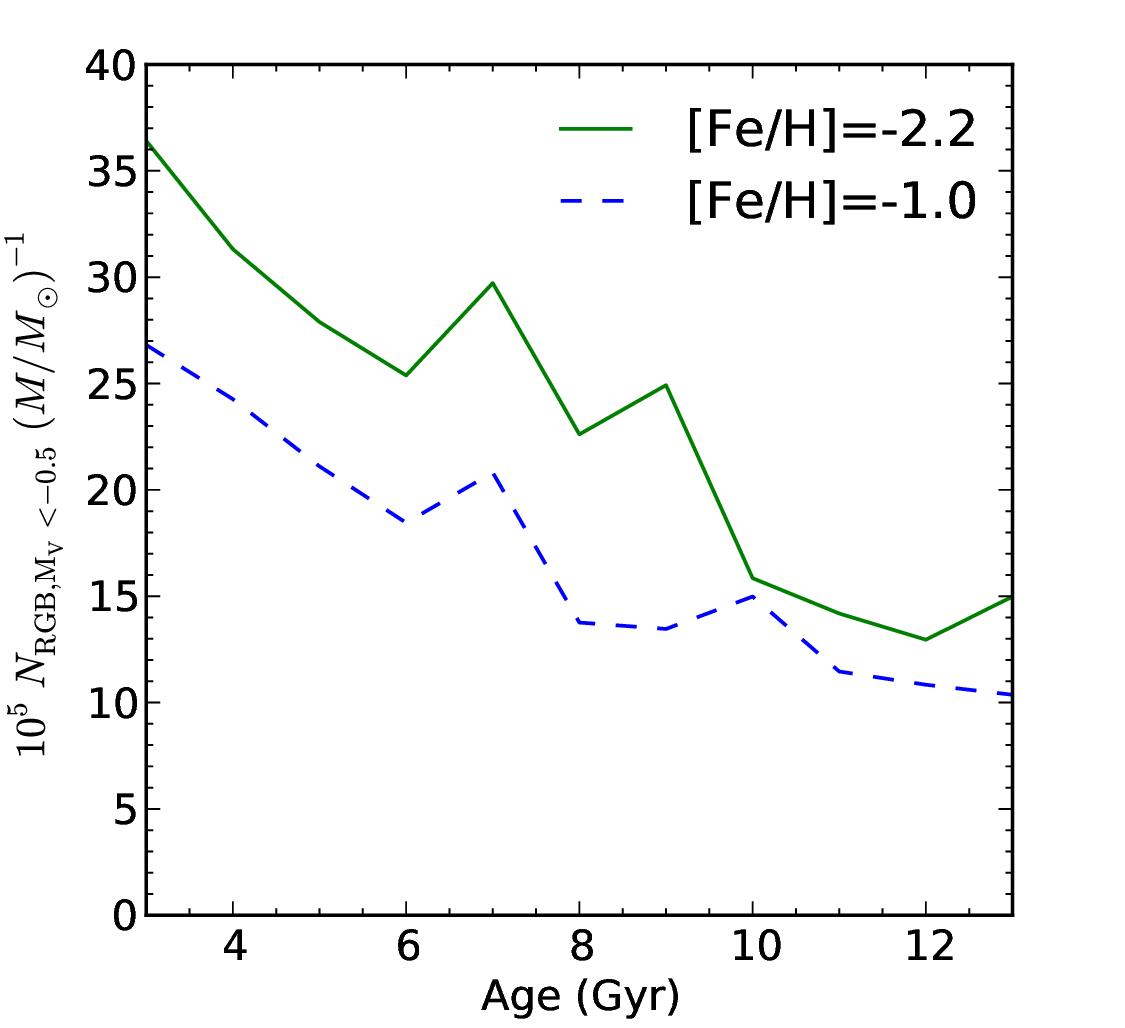}
\caption{\label{fig:nrgb}Number of RGB stars (with $M_V<-0.5$) per $10^5 M_\odot$ formed versus age, based on \citet{Dotter2007} isochrones and assuming a \citet{Salpeter1955} IMF normalized over the range 0.15--100 $M_\odot$.}
\end{figure}

Alternatively, we can scale by  stellar mass. From resolved colour-magnitude diagrams, \citet{Coleman2008} estimated that a total of  $6.1\times10^7 M_\odot$ of stars have formed in Fornax (for a Salpeter IMF extending down to 0.15 $M_\odot$). 
A somewhat smaller mass of $4.3\times10^7 M_\odot$ within 0\fdg8 was found by \citet{DeBoer2012}.
Although the star formation history is complex and extends up to a few 100 Myr ago, the vast majority of the stars formed more than 2--3 Gyr ago, and would therefore contribute to the RGB star population on which the Battaglia et al.\ metallicities are based.
Converting the observed numbers of RGB stars per metallicity bin to  stellar \emph{mass} per bin is, however, slightly non-trivial, as this conversion depends on both age and metallicity. Of interest here is the number of RGB stars that are sufficiently bright to be included in the \citet{Battaglia2006} sample. The faintest stars in their sample are slightly fainter than $V=20$, corresponding to a limit of $M_V\approx-0.5$.
From stellar model tracks  \citep{Dotter2007}, we find that the amount of time an RGB star spends at $M_V<-0.5$ is very nearly independent of age, but decreases with increasing metallicity from about 25 Myr at ${\rm [Fe/H]}=-2$ to about 20 Myr at ${\rm [Fe/H]}=-1$. If there are no age differences, low-metallicity stars would then be somewhat \emph{over-represented} in Fig.~\ref{fig:zglob} relative to the actual amount of mass present at low metallicities. To find the relation between number of RGB stars per unit stellar mass formed and \emph{age}, the RGB lifetime must be multiplied by the specific evolutionary flux, essentially the product of the IMF (evaluated at the turn-off mass) and the rate at which the turn-off mass decreases \citep{Renzini1986}. The net result is that the number of RGB stars per unit mass decreases by roughly a factor of two over the time interval 3--12 Gyr at fixed metallicity (Fig.~\ref{fig:nrgb}). This would then cause \emph{older} stars to be \emph{underrepresented} in Fig.~\ref{fig:zglob}, counteracting the trend with metallicity if there is an age-metallicity relation \citep{Battaglia2006,Coleman2008}. If the bulk of the metal-rich stellar population is as young as 3--4 Gyr and the metal-poor population is very old ($>10$ Gyr), then Fig.~\ref{fig:nrgb} suggests that we will underestimate the mass of the metal-poor population by up to $\sim60$\%.
The actual star formation history has probably been somewhat more continuous over the range $\sim$3--10 Gyr \citep{Battaglia2006,Coleman2008,DeBoer2012}, in which case Fig.~\ref{fig:zglob} would more closely represent the true metallicity distribution by mass.
From similar considerations, we find that the number of RGB stars per unit integrated \emph{luminosity} varies much less with age, while the metallicity dependence is similar to that in Fig.~\ref{fig:nrgb}.  We may then have \emph{overestimated} the fraction of light coming from metal-poor stars by  $\sim$10--20\%.

Scaling the \citet{Coleman2008} mass estimate by 5\%, we then find that $3.1\times10^6 M_\odot$ formed in stars with ${\rm [Fe/H]}<-2$.  About 35\% of the stellar mass will have been lost due to stellar evolution over a Hubble time \citep[e.g.][]{Lamers2005}, leaving $\sim2\times10^6 M_\odot$ in metal-poor stars today.
The combined masses of the four metal-poor GCs are $\sim1.0\times10^6 M_\odot$ assuming an average $M/L_V=3.5$ \citep{Dubath1992,Larsen2012}. From this we find that these clusters have a combined mass of about 1/2  of the mass in metal-poor field stars, and that about 1 in 3 stars with ${\rm [Fe/H]} <-2$ belongs to one of the four metal-poor GCs. If we correct for age effects, as discussed above, this ratio might decrease to 1:4 or 1:5. These estimates are very similar to that based on the luminosities. 
 
These are evidently rather rough estimates. The total mass of Fornax is uncertain, and the  assumption that the metal-poor stars are smoothly distributed may not be valid if they formed in a few discrete events associated with the GCs \citep{Penarrubia2009a}. Note that the FLAMES data preferentially sample fields close to, or including the GCs. 
Also, the dynamically based $M/L_V$ ratio used for the GCs is higher than the $M/L_V\approx2$ predicted by SSP models for a standard IMF \citep[e.g.][]{Strader2011}.
Further, the \ion{Ca}{ii} triplet-based metallicity measurements may not be on exactly the same scale as our high-dispersion spectroscopy of the GCs, although this mainly affects the shape of the metallicity distribution for $\mathrm{[Fe/H]}<-2$ \citep{Starkenburg2010}.
Nevertheless, it is clear that \emph{a staggeringly large fraction of the most metal-poor stars in Fornax belong to the four metal-poor GCs}.  
For comparison, the Milky Way GC system has a combined mass of $\sim2.8\times10^7 M_\odot$ \citep{Kruijssen2009}, or about 2\% of the mass of the stellar halo if a total mass of $\sim10^9 M_\odot$ is assumed for the latter \citep[e.g.][]{Suntzeff1991} and we take into account that only $\sim2/3$ of the Galactic GC system is associated with the halo. 

\section{GC formation and self-enrichment}

The fact that, after a Hubble time of dynamical evolution, a large fraction of the metal-poor stars in the Fornax dSph still belong to the GCs, has important implications for our understanding of the formation and dynamical evolution of GCs. 

First, there is little room for dissolution of any additional clusters formed together with the four extant metal-poor objects. Each of these four surviving metal-poor GCs has likely lost several times $10^5 M_\odot$ due to evaporation over a Hubble time \citep[e.g.][]{Jordan2007}, so that these combined losses may well amount to over a million $M_\odot$. The implication is then that perhaps half of the most metal-poor stars now present in Fornax may originally have formed in these four  GCs. This might suggest that the Fornax GCs did not form as part of a larger population of star clusters with a power-law mass distribution extending down to low masses. This is reminiscent of the star cluster populations in some present-day star forming dwarf galaxies which host a single, or a few, outstandingly massive clusters, one of the most extreme examples being NGC~1705 \citep{OConnell1994,Ho1996}.

Second, these numbers represent a challenge for self-enrichment scenarios in which GCs were initially much more massive than they are now \citep{DErcole2008,Schaerer2011,Bekki2011}. 
It seems to be ruled out that the Fornax GCs could possibly have been more than a factor of a few more massive initially than they are now. This limit may be even more stringent for Fornax 3, the most massive of the clusters, whose very low metallicity  places it even further out in the tail of the field star metallicity distribution.
Nevertheless,  detailed abundance measurements suggest that the Fornax GCs share the same abundance anomalies seen in Galactic GCs.  
It has been suggested that the second generation may have formed partly from gas that was subsequently accreted \citep{Conroy2011,Pflamm-Altenburg2009}.
However, even this possibility does not easily explain the high ratio of GC to field stars in the Fornax dSph. It would require that each cluster was able to accrete a significant amount of gas with the same composition as the first-generation stars, while little gas of this composition was available for the formation of additional field stars. 

An important assumption in the above discussion is that the field star population observed in Fornax today has not been significantly stripped by the Galactic tidal field. While this has likely affected some dSphs, the stellar component in Fornax itself does not appear to have been affected by tidal stripping  \citep{Penarrubia2009}.
Conversely, it is possible that the GCs did not originate within the Fornax dSph, but  formed in halos that were stripped of stars and dark matter before being accreted onto Fornax.
This has also been suggested as a possible solution to the ``timing problem'', i.e., the fact that the clusters have not yet been dragged into the centre of the galaxy as expected due to dynamical friction \citep{Cole2012}. This scenario is difficult to rule out, but it is unclear why four or five globular clusters would have formed close enough to the Fornax dSph to be captured by it, given that no significant population of  inter-galactic GCs has been found elsewhere in the Local Group. 

\section{Conclusions}

We have used the most recent measurements of metallicities for globular clusters and field stars in the Fornax dwarf spheroidal galaxy to estimate the specific frequency of metal-poor GCs. Four of the GCs have ${\rm [Fe/H]} < -2.0$; associating these GCs with field stars in the same metallicity range we estimate $S_N \approx 400$, by a wide margin the highest known anywhere.
We find that about one quarter of all stars in the Fornax dSph with ${\rm [Fe/H]}<-2$ belong to these four most metal-poor clusters. Accounting for standard dynamical evolution, an even larger fraction of the metal-poor stars must initially have been born in the clusters. This puts tight constraints on any further amount of stars that could have been lost from metal-poor star clusters due to early ``infant weight loss'' or ``infant mortality''.  The extant clusters could at most have been a factor of $\sim4-5$ more massive initially then they are now, and this requires a rather extreme scenario in which no field stars or other clusters of similar metallicity were formed initially.

It would be of significant interest to carry out detailed chemical 
tagging of the most metal-poor stars in Fornax to determine how many of them can be traced back to the GCs.
With sufficient accuracy, it should be possible to detect peaks in the field star metallicity distribution corresponding to stars lost from the GCs. One might also expect to see a significant fraction of metal-poor field stars with peculiar, GC-like abundance patterns. Fornax and other dSph galaxies may be unique laboraties in which  the presence or absence of anomalous abundances of light elements in the field stars might shed light on the origin of multiple stellar populations in globular clusters.

\begin{acknowledgements}
We thank Charlie Conroy, Mark Gieles, Tom Richtler and the  referee for valuable comments and discussion. SSL acknowledges support by an NWO/VIDI grant and JPB acknowledges NSF grant AST-1109878.
\end{acknowledgements}

\bibliographystyle{aa}
\bibliography{libmen.bib}

\end{document}